\documentclass[11pt]{article}
\usepackage{latexsym}
\usepackage{graphicx}
\usepackage{amssymb}
\usepackage{amsmath,enumerate,rotate}

\newcommand{\R}{\mathbb R}

\def\be#1\ee{\begin{equation}#1\end{equation}}
\newcommand{\fer}[1]{(\ref{#1})}

\newcommand{\bq}{\begin{equation}}
\newcommand{\eq}{\end{equation}}

\def\bqa{\begin{eqnarray}}
\def\eqa{\end{eqnarray}}

\def\e{\epsilon}

\newcommand{\bd}{\begin{displaymath}}
\newcommand{\ed}{\end{displaymath}}
\newcommand{\ba}{\begin{eqnarray}}
\newcommand{\ea}{\end{eqnarray}}

\def\R{\mathbb{R}}

\newenvironment{equations}{\equation\aligned}{\endaligned\endequation}

\begin{document}

\title{Wealth distribution and collective knowledge.\\ A Boltzmann approach}

\author{L. Pareschi\thanks{Department of Mathematics and Computer Science, University of Ferrara, via Machiavelli 35, 44121 Ferrara, ITALY. (lorenzo.pareschi@unife.it)} \and
G. Toscani\thanks{Department of Mathematics,
University of Pavia, via Ferrata 1, 27100 Pavia, ITALY. (giuseppe.toscani@unipv.it)} }



\maketitle

\begin{abstract}
We introduce and discuss a nonlinear kinetic equation of Boltzmann type
which describes the influence of knowledge in the evolution of wealth in a system of
agents which interact through the binary trades introduced in \cite{CoPaTo05}. The
trades, which include both saving propensity and the risks of the market, are here
modified in the risk and saving parameters, which now are assumed to depend on the
personal degree of knowledge. The numerical simulations show that the presence of knowledge has the potential to produce a class of wealthy agents and to account for a larger proportion of wealth inequality.
\end{abstract}

\section{Introduction}
In the last two decades, various concepts and techniques of statistical mechanics have been
fruitfully applied to a wide variety of complex extended
systems, physical and non-physical, in an effort to understand their
emergent properties. Economics is, by far, one of
the systems to which methods borrowed from
statistical mechanics for particle systems have been systematically applied \cite{
Ch02,ChaCha00,ChChSt05, CoPaTo05,DY00,Ha02,IKR98,Sl04}. Starting
from  the original idea developed by Angle \cite{Ang, Ang1}, a
variety of both discrete and continuous models for wealth distribution has been proposed and
studied in view of the relation between parameters in the
microscopic rules and the resulting macroscopic statistics
\cite{ChaCha00,CoPaTo05, DY00, GSV, SGD}. The basic assumption in most of these models is that wealth is exchanged among agents by means of binary trades, which represent the microscopic level.  A typical ingredient of trading is a mechanism for saving,  first introduced
in \cite{ChaCha00},  which ensures that agents could exchange at most
a certain fraction of their wealth in each trade event. Moreover,
randomness plays a r\^ole in virtually all available models, taking
into account that many trades are risky, so that the exact amount of
money changing hands is not known a priori.
In most of the models
introduced so far, the trading mechanism leaves the total mean
wealth unchanged. Then, depending on the specific choice of the
saving mechanism and the stochastic nature of the trades, the
studied systems are capable to produce wealth curves with the so-called Pareto power law tails \cite{DMT, MaTo07}. In other words, if $f(v)$ is the probability
density function of agents with wealth $v$
\[
\int_v^{\infty} f(w)\,dw \sim v^{-\mu},
\]
where the exponent $\mu$ characterizes the tail behavior and it is related to the amount of wealth inequality in the system.

Among other approaches, the description of market models via kinetic equations of Boltzmann type is a fertile ground for research. This powerful methodology has been recently extended to cover more sophisticated rules for trading. For example, a description of the behavior of a
stock price has been developed by Cordier, Pareschi and Piatecki in
\cite{CPP}. Also, price theory has been considered by introducing trades in which agents interacts by exchanging goods according to the Edgeworth box strategy \cite{TBD}.
Further, there have been efforts to include
non-microscopic effects, like global taxation (and subsequent
redistribution), in recent works of Guala \cite{Gua}, Pianegonda,
Iglesias, Abramson and Vega \cite{Pia}, Garibaldi, Scalas and
Viarengo \cite{GSV} and Bisi, Spiga and Toscani \cite{BST, To2}.
Despite the high number of studies devoted to the subject, well
documented by various recent review papers \cite{CC09,gup,NPT,Pat,yak,
YR}, most of the kinetic models depend in general only from the wealth
parameter, thus describing a society in which all agents apply the
same rules of change to their wealth, independently of any other
behavioral aspect.

To this common line of thought, the few exceptions need to be properly quoted. Maybe
the most important contribution has been done by Lux and Marchesi \cite{LMa, LMb},
who, to give a basis to the possibility to obtain both booms and crashes of economy,
considered a system composed by two different dynamical classes of traders,
characterized by different behaviours with respect to the trading activity. Their idea
has been subsequently applied in \cite{MD} to construct a kinetic model in which the
personal opinion is responsible of the change of strategy, moving the two groups of
traders considered by Lux and Marchesi from one to the other. It is interesting to
remark that in \cite{MD} these behavioral aspects are justified by resorting to the
prospect theory of Kahneman and Tversky \cite{KTa,KTb}, namely to the analysis of
decision under risk. Other  authors studied the importance of the saving parameter in
the binary trade, by assuming that the same is not a universal one, but it can vary in
a random way \cite{CCM}. For a recent introduction to the kinetic description of most of the above mentioned models, together with their numerical simulation by Monte Carlo methods, we refer to the recent book \cite{PT13}.

Among the various behavioral aspects, knowledge represents an important variable to be
taken into account. Indeed,  knowledge is necessary for an individual to gain autonomy
from its natural, physiological, and social restrictions, leading the population to
improve its life conditions. Hence, it is commonly accepted that knowledge plays a
relevant rule in any activity which could lead to an improvement of the social
conditions, and it is out of doubts that trading of wealth belongs to this set of
activities \cite{HHY}. Knowledge is a consequence of learning, which is usually intended as any
relatively permanent change in behavior which occurs as a result of experience or
practice \cite{Mor}. The way in which knowledge is accumulated is a very complicated
matter, and it is very difficult to model it from a mathematical point of view.
However, in view of its importance in multi-agent phenomena, various aspect of learning
have been recently studied in a number of technical papers (cf. \cite{BMO,Cos,Mar,LMM:13} and
the references therein).
In \cite{HiHa:09, Atlas} the authors introduced the notion of economic complexity, as a measure of the total amount of knowledge present in a country's economy. According to their viewpoint, the more diverse and specialized jobs a country's citizens have, the greater the country's ability to produce complex products that few other countries can produce, making the country more prosperous. At the same time economic complexity generates sensible differences in the social conditions among different countries.
Indeed, it is believed that knowledge accumulation can be the source of a large proportion of wealth inequality \cite{LMM:13}.

In this note, we will introduce a very simple and naive process for
the formation of knowledge, which is described in terms of a linear
interaction with a background, and it is able to take into account
both the acquisition of information  and the process of selection,
which appear to be  natural and universal features. The amount of
knowledge will be quantified in terms of a positive variable. The
knowledge interaction will be coupled with the binary trade of wealth exchange introduced in
\cite{CoPaTo05}, which includes both saving propensity and the risks
of the market. The novelty here is to assume that both the saving propensity and the risky part of
the trade depend on the personal knowledge. A typical and
natural assumption is that knowledge could act on the trade process
both to increase the expected utility, and to reduce the randomness present in the risk. The interaction rules will subsequently be merged, within the
principles of classical kinetic theory,  to derive a nonlinear
Boltzmann-like kinetic equation for the joint evolution of
knowledge and wealth.


For the sake of completeness, a comment on the justification of kinetic models suitable to describe human behaviors is in place. The socio-economic behavior of a (real) population of agents is extremely complex.
Apart from elements from mathematics and economics, a sound description -- if one at all exists --would necessarily need contributions from various other fields, including
psychology. Clearly, the available mathematical models are too simple to even pretend to reflect the real situation. However, the idea to describe evolution of knowledge and wealth in terms of a kinetic equation gives rise to a variety of challenging mathematical problems, both from the theoretical and numerical point of view. In particular, it is remarkable that this class of simple
models possesses such a wide spectrum of possible equilibria (some of which indeed resemble distributions seen in the reality, like the presence of a consistent middle class, as well as the persistency of heavy tails in the steady state distribution).
Moreover, kinetic models are extremely flexible with respect to the introduction of additional effects \cite{PT13}. In this way, the described models should be considered as basic
building blocks, that can easily be combined, adapted and improved.

The paper is organized as follows. In Section \ref{know6} we introduce
and discuss a kinetic model for the formation of knowledge in a multi-agent society. This linear model is based on microscopic interactions with a fixed background, and is such that the density of the population knowledge moves towards a steady distribution which is heavily dependent of the microscopic parameters of the microscopic interactions. Then, the knowledge rule is merged with the binary trade of wealth to obtain a nonlinear kinetic model of Boltzmann-type for the joint density of wealth and knowledge. This part is presented in Section  \ref{model}. Section \ref{FPdes} deals with the derivation of a Fokker--Planck type equation, which models the     formation of the joint density of wealth and knowledge in the so-called \emph{quasi-invariant  limit} procedure. Last, Section \ref{nume} is devoted to various numerical experiments, which allow to recover the steady distribution of wealth and knowledge in the population for various choices of the relevant parameters.

\section{An evolutionary model for the knowledge}\label{know6}

Knowledge is in part inherited from parents, but there is general agreement that the
environment is the main factor influencing it. Experiences that produce know\-led\-ge
cannot be inherited from parents, as is the case for the genome, but rather are
acquired in the course of life from different elements of the environment surrounding
each individual \cite{TB}. The process of learning is very complicated and different
for each individual of a population. Although the overall conditions for all
individuals could be the same, after this process each individual has a different
level of knowledge. Differences in the environment or experiences in life, produce
different levels of knowledge and, as a consequence, a different behavior that depends
on the environment surrounding an individual. Also, it is almost evident that the
personal knowledge is the result of a selection, which allows to retain mostly the
notions that the individuals consider important, and to discard the rest. This basic
aspect of the process of learning has been recently discussed in a convincing way by
Umberto Eco \cite{Eco}, one of the greatest philosophers and contemporary Italian
writers. In his fascinating lecture, Eco outlines the importance of a drastic
selection of the surrounding quantity of information, to maintain a certain degree of
ingenuity. This is particularly true nowadays, where the global access to information
via web gives to each individual the possibility to have a \emph{reservoir} of
infinite capacity from which to extract any type of (useful or not) information.

The previous remarks are at the basis of a suitable description of
the evolution of the distribution of knowledge in a population of
agents by means of microscopic interactions with a fixed background.
In this picture each variation of knowledge is interpreted as an
interaction where a fraction of the knowledge of the individual is
lost by virtue of the selection, while at the same time the external
background (the surrounding environment) can move a certain amount
of its knowledge to the individual. If we quantify the knowledge of
the individual in terms of a scalar parameter $x$, ranging from zero
to infinity, and by $z \ge 0$ the knowledge achieved from the
background in a single interaction, the new amount of knowledge can
be computed as
 \be\label{k1}
 x^* = (1-\lambda(x))x + \lambda_B(x) z + \kappa x.
 \ee
In \fer{k1} the functions $\lambda= \lambda(x)$ and $\lambda_B=
\lambda_B(x)$ quantify, respectively, the amounts of selection and
external learning, while $\kappa$ is a random parameter which takes
into account the possible unpredictable modifications of the
knowledge process. We will in general fix the mean value of $\kappa$
equal to zero. Since some selection is always present, and at the same time it can not exceed a certain amount
of the total knowledge, it is assumed that $\lambda_- \le \lambda(x) \le
\lambda_+$, where $\lambda_- >0$, and $\lambda_+ < 1$. Likewise, it is reasonable to assume an upper bound for the acquisition of knowledge from the background. Then,  $0 \le \lambda_B(x)\le \bar\lambda$, where $\bar\lambda <1$.  Lastly, the random part is
chosen to satisfy the lower bound $\kappa \ge - (1-\lambda_+)$. By
these assumptions, it is assured that the post-interaction value
 $x^*$ of the knowledge is nonnegative.

Let $C(z)$, $z \ge 0$ denote the probability distribution of
knowledge of the (fixed) background. It is reasonable to assume that
$C(z)$ has a bounded mean, so that
 \be\label{ba1}
 \int_{\R_+} C(z) \, dz = 1; \quad \int_{\R_+} z\,C(z) \, dz = M
  \ee
We note that the distribution of the background will induce a
certain policy of acquisition of knowledge. For example, let us assume that the
background is a random variable uniformly distributed on the
interval $(0, a)$, where $a > 0$ is a fixed constant. If we choose
for simplicity $\lambda(x) = \lambda_B(x)= \bar\lambda$, and the individual has a degree of knowledge $x>a$, in absence
of randomness the interaction  will always produce a value $x^* \le
x$, namely a partial decrease of knowledge. In this case, in fact,
the process of selection in an individual with high knowledge can
not be restored by interaction with the environment.

The study of the
time-evolution of the distribution of knowledge consequent to
interactions of type \fer{k1} among individuals can be obtained by
resorting to kinetic collision-like models \cite{PT13}. Let $F=
F(x,t)$ the density of agents which at time $t >0$ are represented by their
knowledge $x \in \R_+$. Then, the time
evolution of $F(x, t)$  obeys to a
Boltzmann-like equation. This equation can be fruitfully be written
in weak form. It corresponds to say that the solution $F(x,t)$
satisfies, for all smooth functions $\varphi(x)$ (the observable quantities)
 \be
  \label{kine-w}
  \frac{d}{dt}\int_{\R_+}F(x,t)\varphi(x)\,dx  =
  \Big \langle \int_{\R_+^2} \bigl( \varphi(x^*)-\varphi(x) \bigr) F(x,t)C(z)
\,dx\,dz \Big \rangle.
 \ee
In \fer{kine-w}  the post-interaction knowledge $x^*$ is given by  \fer{k1}. As
usual, $\langle \cdot \rangle$ represents mathematical expectation. Here expectation takes into account the presence of the random parameter $\kappa$ in \fer{k1}.

The meaning of the kinetic equation \fer{kine-w} is clear. At any positive time $t >0$, the variation in time of the distribution of knowledge (the left-hand side) is due to the change in knowledge resulting from interactions of type \fer{k1} that the system of agents has at time $t$ with the environment. This change is measured by the interaction operator at the right-hand side.

It is immediate to show that equation \fer{kine-w} preserves the
total mass, so that $F(x,t)$, $t >0$, remains a probability density if it is so
initially. By choosing $\varphi(x) = x$ we recover the evolution of
the mean knowledge $M_K(t)$ of the agents system. The mean knowledge satisfies the equation
 \be\label{m2}
 \frac{dM_K(t)}{dt} = - \int_{\R_+} x \lambda(x)F(x,t) \, dx +
 M \int_{\R_+}  \lambda_B(x)F(x,t) \, dx,
 \ee
which in general it is not explicitly solvable, unless the functions $\lambda(x)$ and $\lambda_B(x)$ are assumed to be constant.
However, since $\lambda(x) \ge
\lambda_-$, while $ \lambda_B(x) \le \bar\lambda$, the mean value
always satisfies the differential inequality
 \be\label{m3}
 \frac{dM_K(t)}{dt} \le - \lambda_- M_K(t) + \bar\lambda M,
 \ee
which guarantees that the mean knowledge of the system will never
exceed the (finite) value $M_{max}$ given by
 \[
M_{max} = \frac{ \bar\lambda}{\lambda_-} M.
 \]
If  $\lambda(x)= \lambda $ and $\lambda_B(x)=
\lambda_B$ are constant, equation \fer{m2}
becomes
 \be\label{m5}
 \frac{dM_K(t)}{dt} = - \lambda M_K(t) +
  \lambda_B M.
 \ee
In this case, the linear differential equation can be solved, and
 \be\label{ex1}
M_K(t) = M_K(0) e^{-\lambda t} + \frac{\lambda_B M}{\lambda} \left(
 1- e^{-\lambda t} \right).
 \ee
Formula \fer{ex1} shows that the mean knowledge converges exponentially to its limit value $\lambda_B M/\lambda$.

\section{A Boltzmann model for wealth and knowledge}\label{model}
\setcounter{equation}{0}

In this section, we will join the evolutionary kinetic model for
knowledge with a kinetic model for wealth distribution we introduced
with Cordier in 2005 \cite{CoPaTo05}. This model belongs to a class
of models in which agents are indistinguishable. In most of these
models \cite{DMT, DMT1} an agent's \emph{state} at any instant of
time $t\geq0$ is completely characterized by his current wealth
$w\geq0$. When two agents encounter in a trade, their {\em pre-trade
wealths\/} $v$, $w$ change into the {\em post-trade wealths\/}
$v^*$, $w^*$ according to the rule \cite{Ch02, ChaCha00,ChChSt05}
\begin{equation}
  \label{eq.trules}
  v^* = p_1 v + q_1 w, \quad w^* = q_2 v + p_2 w.
\end{equation}
The {\em interaction coefficients\/} $p_i$ and $q_i$ are
non-negative random variables. While $q_1$ denotes the fraction of
the second agent's wealth transferred to the first agent, the
difference $p_1-q_2$ is the relative gain (or loss) of wealth of the
first agent due to market risks. It is usually assumed that $p_i$
and $q_i$ have fixed laws, which are independent of $v$ and $w$, and
of time. This means that the amount of wealth an agent contributes
to a trade is (on the average) proportional to the respective
agent's wealth.

Let $f(v,t)$ be the density of agents which at time $t >0$ are represented by their
wealth $v \in \R_+$. As explained in Section \ref{know6},  the
time-evolution of the distribution of wealth, say $f(v,t)$, consequent to binary
interactions of type \fer{eq.trules} among agents of the system, is obtained by
resorting to kinetic collision-like models \cite{CoPaTo05, PT13}. The time
evolution of the distribution of wealth  obeys here to a bilinear
Boltzmann-like equation, that in weak form reads
\be
  \label{kine-ww}
  \frac{d}{dt}\int_{\R_+}f(v,t)\varphi(v)\,dv  = \frac 12
  \Big \langle \int_{\R_+^2} \bigl(\varphi(v^*)+ \varphi(w^*)-\varphi(v)-\varphi(w) \bigr) f(v,t)f(w,t)
\,dv\,dw \Big \rangle.
 \ee
In \fer{kine-ww}  the post-interaction wealths $v^*$ and $w^*$  are given by  \fer{eq.trules}. Also in this case, $\langle \cdot \rangle$ represents mathematical expectation, and expectation takes into account the fact that the {\em interaction coefficients\/} $p_i$ and $q_i$ are
non-negative random variables.

In \cite{CoPaTo05}  the trade has been modelled to include the idea that wealth changes hands for a specific reason: one agent intends to {\em invest\/}\nobreakspace his wealth in some asset,
property etc.\ in possession of his trade partner. Typically, such investments bear
some risk, and either provide the buyer with some additional wealth, or lead to the
loss of wealth in a non-deterministic way. An easy realization of this idea
\cite{MaTo07} consists in coupling saving propensity  with some {\em risky
investment\/} that yields an immediate gain or loss proportional to the current wealth
of the investing agent
\begin{equation}
  \label{eq.cpt}
  v^* = \Bigl(1-\gamma+\eta_1\Bigr)v + \gamma w, \quad
  w^* = \Bigl(1-\gamma +\eta_2\Bigr)w + \gamma v,
\end{equation}
where $0 <\gamma <1$ is the parameter which identifies the saving propensity. In this case
 \be\label{cpt1}
   p_i = 1-\gamma + \eta_i , \quad q_i = \gamma \quad (i=1,2).
  \ee
The coefficients $\eta_1,\eta_2$ are random parameters, which are
independent of $v$ and $w$, and distributed so that always
$v^*,\,w^*\geq  0$, i.e.\ $\eta_1,\,\eta_2\geq-\gamma$. Unless these
random variables are centered, i.e.\
$\langle\eta_1\rangle=\langle\eta_2\rangle=0$, it is immediately
seen that the mean wealth is not preserved, but it increases or
decreases exponentially (see the computations in \cite{CoPaTo05}).
For centered $\eta_i$,
\begin{equation}\label{con}
  \langle v^* + w^* \rangle = (1+\langle\eta_1\rangle) v
  + (1+\langle\eta_2\rangle) w = v + w ,
\end{equation}
implying conservation of the average wealth. Various specific
choices for the $\eta_i$ have been discussed \cite{MaTo07}. The
easiest one leading to interesting results is $\eta_i=\pm r$, where
each sign comes with probability $1/2$. The factor
$r\in(0,\gamma)$ should be understood as the {\em intrinsic
risk\/} of the market: it quantifies the fraction of wealth agents
are willing to gamble on. Within this choice, one can display the
various regimes for the steady state of wealth in dependence of
$\gamma$ and $r$, which follow from numerical evaluation. In the
zone corresponding to low market risk, the wealth distribution shows
again \emph{socialistic} behavior with slim tails. Increasing the
risk, one falls into \emph{capitalistic}, where the wealth
distribution displays the desired Pareto tail. A minimum of saving
($\gamma>1/2$) is necessary for this passage; this is expected since
if wealth is spent too quickly after earning, agents cannot
accumulate enough to become rich. Inside the capitalistic zone , the
Pareto index decreases from $+\infty$ at the border with
\emph{socialist} zone  to unity. Finally, one can obtain a steady
wealth distribution which is a Dirac delta located at zero. Both
risk and saving propensity are so high that a marginal number of
individuals manages to monopolize all of the society's wealth. In
the long-time limit, these few agents become infinitely rich,
leaving all other agents truly pauper.

The analysis in \cite{MaTo07} essentially show that the microscopic interaction \fer{eq.cpt} considered in Cordier,
Pareschi and Toscani model (briefly CPT model) are such that the kinetic equation \fer{kine-ww} is able to describe all interesting behaviors of wealth distribution in a
multiagent society. In its original formulation, both the saving and
the risk were described in terms of the universal constant $\gamma$
and of the universal random parameters $\eta_1,\eta_2$. Suppose now
that these quantities in the trade could depend of the personal
knowledge of the agent. For example, one reasonable assumption would
be that an individual uses his personal knowledge to reduce the risk
connected to trading. Also, knowledge could be used to have a
favorable outcome from the trade.  Under these assumptions, one is led
to modify the binary trade to include the effect of knowledge into
\fer{eq.cpt}. Given two agents $A$ and $B$ characterized by the pair
$(x,v)$ (respectively $(y,w)$) of knowledge and wealth, the new
binary trade between $A$ and $B$ now reads
\begin{equations}
  \label{eq.cpt2}
  &v^* = \Bigl(1-\Psi(x)\gamma +\Phi(x)\eta_1\Bigr)v +  \Psi(y)\gamma w,
   \\
 &w^* = \Bigl(1- \Psi(y)\gamma +\Phi(y)\eta_2\Bigr)w +
  \Psi(x)\gamma v.
\end{equations}
In \fer{eq.cpt2} the personal saving propensity and risk perception
of the agents are contained into the functions $\Psi =\Psi(x)$ and
$\Phi= \Phi(x)$. In this way, the outcome of binary trade results
from a combined effect of (personal) saving propensity, knowledge
and wealth. Among other possibilities, one reasonable choice is to
fix the functions $\Psi(\cdot)$ and $\Phi(\cdot)$ as non-increasing
functions. This reflects the idea that the knowledge could be
fruitfully employed both to improve the result of the outcome and to
reduce the risks. For example, $\Psi(x) = (1+x)^{-\alpha}$ and
$\Phi(x) = (1+x)^{-\beta}$, with $\alpha, \beta >0$ could be one
possible choice.

It is interesting to remark that, even in presence of the personal
knowledge (through the functions $\Psi$ and $\Phi$), the
trade \fer{eq.cpt2} remains \emph{conservative in the mean}, that is
 \be\label{cons}
\langle v^*(x,y,v,w) + w^*(x,y,v,w) \rangle = v + w,
 \ee
like in the original CPT model (cf. equation \fer{con}.

Assuming the binary trade \fer{eq.cpt2} as the microscopic binary exchange of wealth in the system of agents, the joint evolution of wealth and knowledge  is described in terms of the density $f=
f(x,v,t)$ of agents which at time $t >0$ are represented by their
knowledge $x \in \R_+$ and wealth $v\in \R_+$. The evolution in
time of the density $f$ is described by the following
kinetic equation (in weak form) \cite{PT13}
 \[
\frac{d}{dt}\int_{\R_+^2}\varphi(x,v) f(x,v,t)\,dx\, dv  = \frac 12
\Big \langle \int_{\R_+^5} \bigl( \varphi(x^*,v^*) +
\varphi(y^*,w^*)
 \]
 \be
  \label{kine-xv}
 -\varphi(x,v) - \varphi(y,w) \bigr) f(x,v,t)f(y,w,t)C(z)
\,dx\,dy\,dz\,dv\, dw \Big \rangle,
 \ee
In \fer{kine-xv} the pairs $(x^*, v^*)$ and  $(y^*, w^*)$ are
obtained from the pairs $(x,v)$ and $(y,w)$  by \fer{k1} and
\fer{eq.cpt2}.  Note that, by choosing $\varphi$ independent of $v$,
that is $\varphi=\varphi(x)$, equation \fer{kine-xv} reduces to the
equation \fer{kine-w} for the marginal density of knowledge
$F(x,t)$. In view of the particular interaction rules, the solution
to equation satisfies some important conservation properties. Let us
define by $M_w(t)$ the mean wealth present in the system at time $t
>0$, that is
 \be\label{mw}
M_W(t) = \int_{\R_+^2} v f(x,v,t) \, dv\, dx.
 \ee
Then, since interactions of type \fer{eq.cpt2} preserve (in the
mean) the total wealth in the single trade,  by
choosing $\varphi(x,v) = v$ it follows that $M_W(t)$ is preserved in
time. Together with mass conservation, this is the only preserved
quantity. In reason of the complicated structure of the binary trade \fer{eq.cpt2}, which depends on the knowledge variable in a nonlinear way, the analytical study of the kinetic equation \fer{kine-xv} appears extremely difficult. For this reason, in the remaining of this paper we will resort to some simplification of the system description, and to numerical experiments.

\section{Fokker-Planck description}\label{FPdes}
\setcounter{equation}{0}

As it is usual in kinetic theory, particular asymptotics of the Boltzmann-type equation result in simplified models, generally of Fokker-Planck type, for which the study of the theoretical properties is often easier \cite{CoPaTo05, To06}.
In order to describe the asymptotic process, let us discuss in some details
the evolution equation for the mean knowledge, given by \fer{m5}. For simplicity, and without loss of generality, let us assume $\lambda$ and $ \lambda_B$ constant.
Given a small parameter $\e$, the scaling
 \be\label{scal}
\lambda \to \e\lambda,\quad \lambda_B \to \e \lambda_B, \quad \kappa
\to \sqrt\e \kappa
 \ee
 is such that the mean value $M_K(t)$ satisfies
 \[
\frac{dM_K(t)}{dt} = -\e\left( \lambda M_K(t) -
  \lambda_B M\right).
 \]
If we set $\tau = \e t$, $F_\e(x,\tau) = F(x,t)$, then
 \[
M_k(\tau) = \int_{\R_+} x F_\e(x,\tau) \, dx = \int_{\R_+} x
F(x,\tau) \, dx = M_K(t),
 \]
and the mean value of the density $F_\e(x,\tau)$ solves
  \be\label{meanscal}
 \frac{dM_K(\tau)}{d\tau} = -  \lambda M_K(\tau) +
  \lambda_B M.
 \ee
Note that equation \fer{meanscal} does not depend explicitly on the
scaling parameter $\e$. In other words, we can reduce in each
interaction the variation of knowledge, waiting enough time to get
the same law for the mean value of the knowledge density.

We can consequently investigate the situation in which most of the
interactions produce a very small variation of knowledge ($\e \to
0$), while at the same time  the evolution of the knowledge density
is such that \fer{meanscal} remains unchanged. We will call this
limit quasi-invariant knowledge limit. Analogous procedures have
been successfully applied to kinetic models in economics
\cite{CoPaTo05} and opinion formation \cite{To06}.

The same strategy applies to the microscopic wealth exchange, if we scale the parameters quantifying the propensity and risk in \fer{eq.cpt2}
 \be\label{scal2}
 \gamma \to
\e\gamma,\quad \eta_i  \to \sqrt\e \eta_i, \quad i =1,2.
 \ee
Let now assume that the centered random variable $\kappa$ has bounded moments at least
of order $n=3$, with $\langle \kappa^2 \rangle = \delta$. Likewise, let us assume that
the centered random variables $\eta_i$, $i=1,2$ are independent and equidistributed,
with bounded moments up to order three, and such that $\langle \eta_i^2 \rangle =
\sigma$, $i=1,2$. Moreover, we assume that $\kappa$ is independent of $\eta_i$,
$i=1,2$.

Using the previous properties of the random quantities $\kappa, \eta_1, \eta_2$,
equations \fer{k1} for the knowledge variable, and  \fer{eq.cpt2} for the wealth give
 \begin{equations}
  \label{var1}
  & \langle x^* -x \rangle =  \lambda_B(x) z - \lambda(x)x = D(x,z),
   \\
 & \langle v^* -v \rangle = \gamma \Bigl( \Psi(y)w -\Psi(x)v \Bigr)
  = E(x,y,v,w),
\end{equations}
and
 \begin{equations}
  \label{var2}
  & \langle (x^* -x)^2 \rangle = D(x,z)^2 + \delta x^2,
   \\
 & \langle (v^* -v)^2 \rangle = E (x,y,v,w)^2 + \sigma \Phi^2(x)v^2,
  \\
  & \langle (x^* -x)(v^*-v) \rangle = D(x,z) E (x,y,v,w).
\end{equations}
Hence, by expanding the smooth function $\varphi(x^*,v^*)$ in Taylor series up to
order two, we obtain
\begin{align}
  \nonumber
  &{\langle \varphi(x^* ,v^*)-\varphi(x ,v) \rangle} = \\
   & D(x,z)\frac{\partial \varphi}{\partial x} +E(x,y,v,w)\frac{\partial \varphi}{\partial v} +
  \frac 12{\delta x^2} \frac{\partial^2 \varphi}{\partial x^2} + \frac 12\sigma{\Phi^2(x)} v^2 \frac{\partial^2 \varphi}{\partial
   v^2} + \\
   \nonumber
   & \frac 12\left( D(x,z)^2 \frac{\partial^2\varphi}{\partial x^2} + E (x,y,v,w)^2 \frac{\partial^2\varphi}{\partial
   v^2} + D(x,z) E (x,y,v,w)\frac{\partial^2\varphi}{\partial x \partial
   v}\right) + {R(x,y,v,w)}.
\end{align}
Clearly, $R(x,y,v,w)$ denotes the remainder of the Taylor expansion.

Let us consider now the situation in which both the scaling \fer{scal} and \fer{scal2}
are applied. Then
 \be\label{scal3}
D(x,z) \to \e D(x,z), \quad E (x,y,v,w) \to \e E (x,y,v,w), \quad \delta \to \e
\delta, \quad \sigma \to \e \sigma.
 \ee
Consequently,
 \[
D(x,z)\frac{\partial \varphi}{\partial x} +E(x,y,v,w)\frac{\partial \varphi}{\partial
v} +  \frac 12{\delta x^2} \frac{\partial^2 \varphi}{\partial x^2} + \frac
12\sigma{\Phi^2(x)} v^2 \frac{\partial^2 \varphi}{\partial   v^2} \to
 \]
 \[
\e\left( D(x,z)\frac{\partial \varphi}{\partial x} +E(x,y,v,w)\frac{\partial
\varphi}{\partial v} +
  \frac 12{\delta x^2} \frac{\partial^2 \varphi}{\partial x^2} + \frac 12 \sigma{\Phi^2(x)} v^2 \frac{\partial^2 \varphi}{\partial
   v^2}\right),
 \]
while
 \[
D(x,z)^2 \frac{\partial^2\varphi}{\partial x^2} + E (x,y,v,w)^2
\frac{\partial^2\varphi}{\partial
   v^2} + D(x,z) E (x,y,v,w)\frac{\partial^2\varphi}{\partial x \partial
   v} \to
 \]
 \[
\e^2 \left( D(x,z)^2 \frac{\partial^2\varphi}{\partial x^2} + E (x,y,v,w)^2
\frac{\partial^2\varphi}{\partial
   v^2} + D(x,z) E (x,y,v,w)\frac{\partial^2\varphi}{\partial x \partial
   v} \right)
 \]
Within the same scaling
 \[
 R(x,y,v,w) \to R_\e(x,y,v,w).
 \]
We remark that, by construction,  the scaled remainder $R_\e(x,y,v,w)$ depends in a
multiplicative way on higher moments of the random variables $\sqrt\e \kappa$ and
$\sqrt\e \eta_i$, $i=1,2$, so that $ R_\e(x,y,v,w)/ \e \ll 1$ for $\e\ll 1$ (cf. the
discussion in \cite{CPP, To06}, where similar computations have been done explicitly).
Identical formulas hold for the differences $\langle y^* -y\rangle$ and $\langle w^*
-w\rangle$, which are obtained simply exchanging $x$ with $y$, $v$ with $w$ and viceversa in  \fer{var1} and in later. If we now
set $\tau = {\e} t$, and for any given $\e$ we define $f(x,v,t) = h_\e(x,v,\tau)$, we
obtain that $h_\e(x,v,\tau)$ satisfies
 \begin{align}
 \label{fp1}
  & \frac{d}{d\tau} \int_{\R^2_+}  \varphi(x, v) h_\e(x,v,\tau)dx dv  = \\
 \nonumber
   & \int_{\R^2_+} \left[ D(x)\frac{\partial \varphi}{\partial x} + E (x,v,t) \frac{\partial \varphi}{\partial v}
    + \frac 12{\delta} x^2 \frac{\partial^2 \varphi}{\partial x^2} +  \frac 12{\sigma}\Phi^2(x) v^2 \frac{\partial^2 \varphi}{\partial
   v^2}\right] h_\e(x,v,\tau) dx dv +
   {\tilde R_\e}(\varphi) .
\end{align}
In \fer{fp1} we denoted
 \begin{equations}
  \label{var3}
  & D(x) = \int_{\R_+} D(x,z) C(z) \, dz =  \lambda_B(x) M -\lambda(x) x,
   \\
 & E(x,v, t) = \int_{\R_+^2} E (x,y,v,w) h_\e(y,w,\tau)\, dy dw ,
  \\
  & \tilde R_\e(\varphi) = \frac\e{2}  \int_{\R^2_+} \int_{\R^2_+} \left[\left( D(x,z)^2 \frac{\partial^2\varphi}{\partial x^2} + E (x,y,v,w)^2
\frac{\partial^2\varphi}{\partial v^2
  } + D(x) E (x,y,v,w)\frac{\partial^2\varphi}{\partial x \partial
   v} \right)\right.\\
   &\null\qquad\qquad\qquad\qquad
   + \left.\frac 1{\e^2} R_\e(x,y,v,w)\right]h_\e(x,v,\tau)h_\e(y,w,\tau)\, dx \, dy\,dv\, dw.
\end{equations}
When the remainder converges to zero as $\e \to 0$,  the density $h_\e(v,w,\tau)$
tends towards $h(x,v,\tau)$ satisfying
\begin{equations}\label{ww1}
  & \frac{d}{d\tau} \int_{\R^2_+}   \varphi(x, v) h(x,v,\tau)\, dx dv = \\
   & \int_{\R^2_+}  \left[\frac 12{\delta} x^2 \frac{\partial^2 \varphi}{\partial x^2} +  \frac
12{\sigma}\Phi^2(x) v^2 \frac{\partial^2 \varphi}{\partial
   v^2} + D(x)\frac{\partial \varphi}{\partial x} + E (x,v,t) \frac{\partial \varphi}{\partial w}\right] h(x,v,\tau) dx dv
\end{equations}
Equation \fer{ww1} is the weak form of the Fokker-Planck equation
 \be\label{FPa}
 \frac{\partial h}{\partial\tau} =  \frac 12{\delta} \frac{\partial^2 (x^2 h)}{\partial x^2}+ \frac
{\sigma\Phi^2(x)}2 \frac{\partial^2 ( v^2h)}{\partial
   v^2} + \frac{\partial (D(x)h) }{\partial x} +
 \frac{\partial (E (x,v,t) h )}{\partial v}
  \ee
The formal derivation of the Fokker-Planck equation \fer{FPa} can be made rigorous
repeating the analogous computations of \cite{CPP, To06}, which refer to
one-dimensional models. This derivation requires to start with initial data that
possess moments  bounded up to a certain order (typically $2+\epsilon$, with $\epsilon
>0$). Moreover, the smooth function $\varphi$ is required to belong to a suitable space (for example $\varphi \in
C^3_{b}$, the space of functions of bounded support continuous with three derivatives)
\cite{To06}.  It is interesting to remark that the balance $\delta/\lambda =C$
(respectively {$\sigma/\gamma = C$} ) are the right ones which maintain in the limit
equation both the effects of knowledge in terms of the intensity $\lambda$, as well as
the effects of the randomness through the variance $\sigma$ (respectively the effects
of saving in terms of $\gamma$ and risk in terms of $\mu_i$, $i=1,2$). As explained in
\cite{To06} for the case of opinion formation, different balances give in the limit
purely diffusive equations, of purely drift equations.

Returning to the original expressions of the quantities $D$ and $E$ into \fer{FPa} we
finally obtain that $h= h(x,v,\tau)$ solves the Fokker-Planck equation
 \be\label{FP}
 \frac{\partial h}{\partial \tau} = \frac \delta 2 \frac{\partial^2 (x^2 h)}{\partial
 x^2}+ \frac{\sigma\Phi^2(x)}2 \frac{\partial^2 (v^2h)}{\partial v^2} +
 \frac{\partial}{\partial x}\left[ (x \lambda(x)- \lambda_B M)h
 \right] +
  \gamma \frac{\partial}{\partial v}\left[ (\Psi(x)v- M_{W}(t))h
 \right],
 \ee
where we denoted
 \be\label{mm}
M_{W}(t) = \left\langle \int_{\R_+^2} w \Psi(y) h(y,w,t) \,dy \, dw \right\rangle.
 \ee
In the simpler case in which the saving propensity remains a universal constant, so
that $\Psi(y) = 1$, the drift term in the Fokker-Planck equation \fer{FP} simplifies,
and, by resorting to the conservation of the mean wealth, we obtain that the density
$h= h(x,v,\tau)$ solves the equation
 \be\label{FP2}
 \frac{\partial h}{\partial \tau} = \frac \delta 2 \frac{\partial^2 (x^2 h)}{\partial
 x^2}+ \frac{\sigma\Phi^2(x)}2 \frac{\partial^2 (v^2 h)}{\partial v^2} +
 \frac{\partial}{\partial x}\left[ (x \lambda(x)- \lambda_B M)h
 \right] + \gamma \frac{\partial}{\partial v}\left[ (v- M_W)h
 \right],
 \ee
where now $M_W$ represents the (constant) value of the quantity in \fer{mm}.

Despite its apparent simplicity (with respect to the kinetic model \fer{kine-xv}), the
analytic description of the steady state of the Fokker--Planck equation \fer{FP2} is
difficult. Therefore, we will resort to numerical computations of the solution to both
the Boltzmann and Fokker--Planck equations for the joint density.

\section{Numerical experiments}\label{nume}

This section contains a numerical description of the solutions  to both
the Boltzmann-type equation \fer{kine-xv} and the Fokker-Planck
equations \fer{FP} and \fer{FP2}. For the numerical approximation of the Boltzmann equation we apply a Monte Carlo method, as described in Chapter 4 of \cite{PT13}. If not otherwise stated the kinetic simulation has been performed with $N=10^6$ particles.

The numerical experiments will
help to clarify the role of knowledge in the final distribution of
the wealth density among the agents. It is evident that, thanks to the mean
wealth conservation, the density $f(x,v,t)$ will converge to a
stationary distribution \cite{PT13}. As usual in kinetic theory,
this stationary solution will be reached in an exponentially fast
time.

In what follows we will denote by  $X(t)$ the random process which
represents the knowledge  of the population at time $t>0$. Its
density is given by the solution of equation \fer{kine-w}. Hence
  \[
  F(x, t) \,dx = P(X(t) \in (x, x+dx)).
  \]
We will denote by $\mathcal{F}$ the distribution function of $X(t)$,
that is
 \[
\mathcal{F}(x, t) = P(X(t) < x) = \int_0^x F(y,t) \, dy.
 \]
Likewise, we will denote by $W(t)$ the random process which
represents the wealth of the population at time $t>0$. Given the
 solution of equation \fer{kine-xv}, its density is given by the marginal density
  \[
  G(v, t) \,d v= P(W(t) \in (v, v+dv))= dv \, \int_\R dx f(x,v,t).
  \]
Lastly, $\mathcal{G}$ will denote the distribution function
 \[
\mathcal{G}(v, t) = P(W(t) < v) = \int_0^ v G(w,t) \, dw.
 \]
We will evaluate the marginal distributions $\mathcal{F}$ and
$\mathcal{G}$ for different values of the parameters $\lambda,
\lambda_B$ and different kinds of microscopic wealth interactions.
In this way we will recognize first the role of background and
selection in the distribution of knowledge (differences in tails
etc.), and, second,  the importance of knowledge (through the
personal saving and risk perception) in the distribution of wealth.

Numerical experiments will also report the joint density of wealth
and know\-led\-ge in the agent system. The following numerical tests have been considered.

\begin{figure}[t]
\includegraphics[scale=.38]{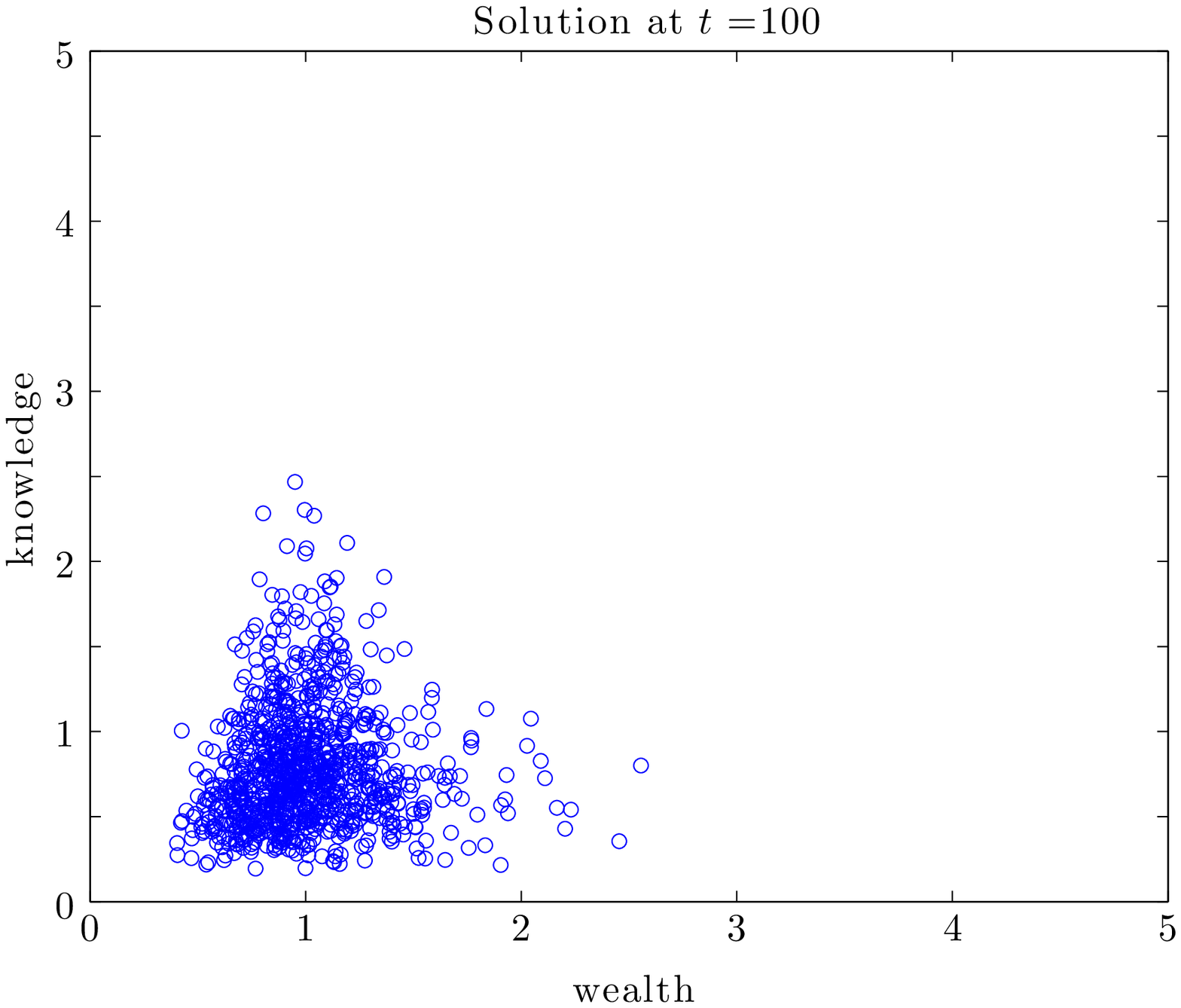}
\includegraphics[scale=.38]{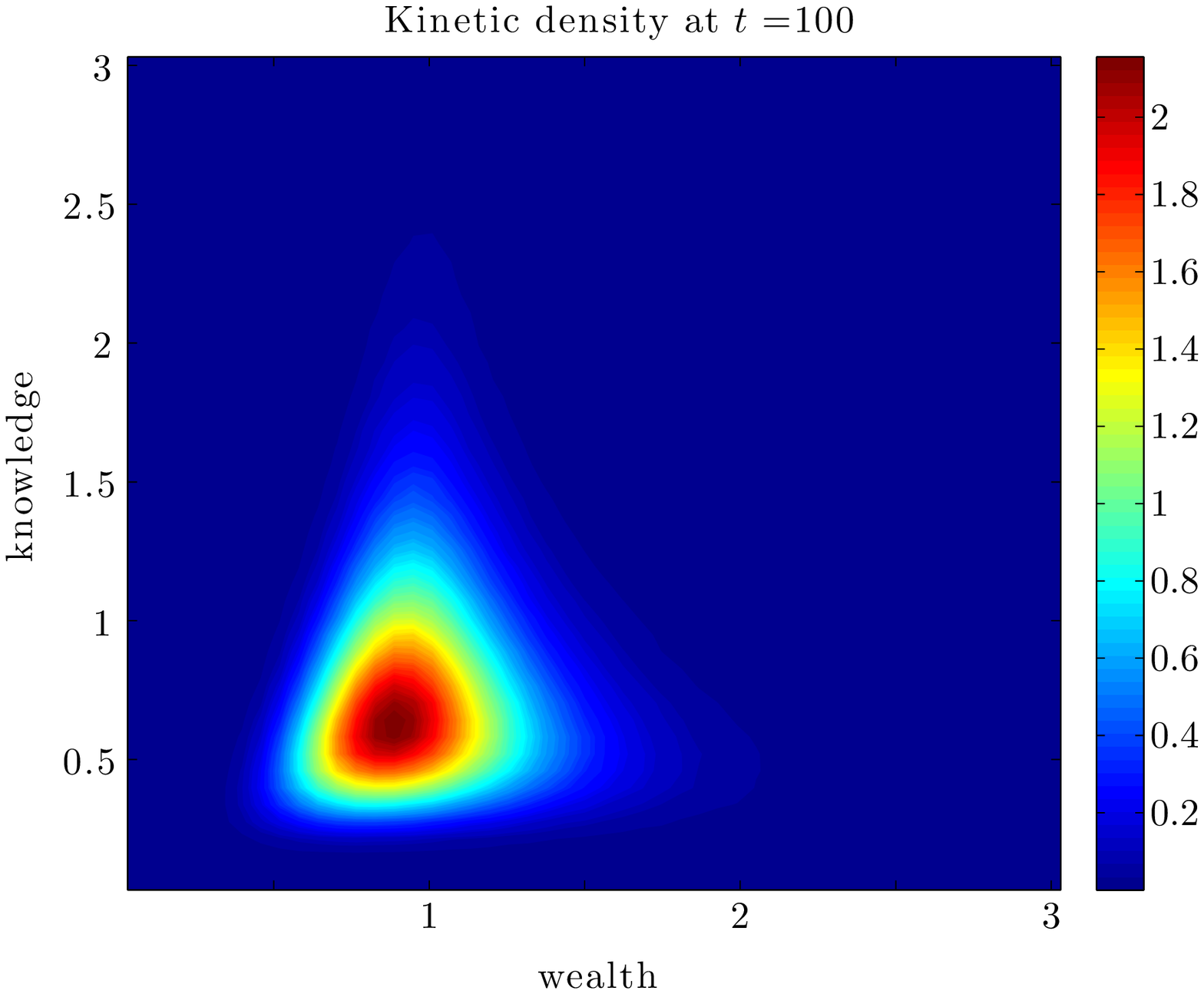}
\caption{Test 1: The particles solution with $N=1000$ particles (left) and the kinetic density (right).}
\label{fg:fig1}
\end{figure}
\begin{figure}
\includegraphics[scale=.38]{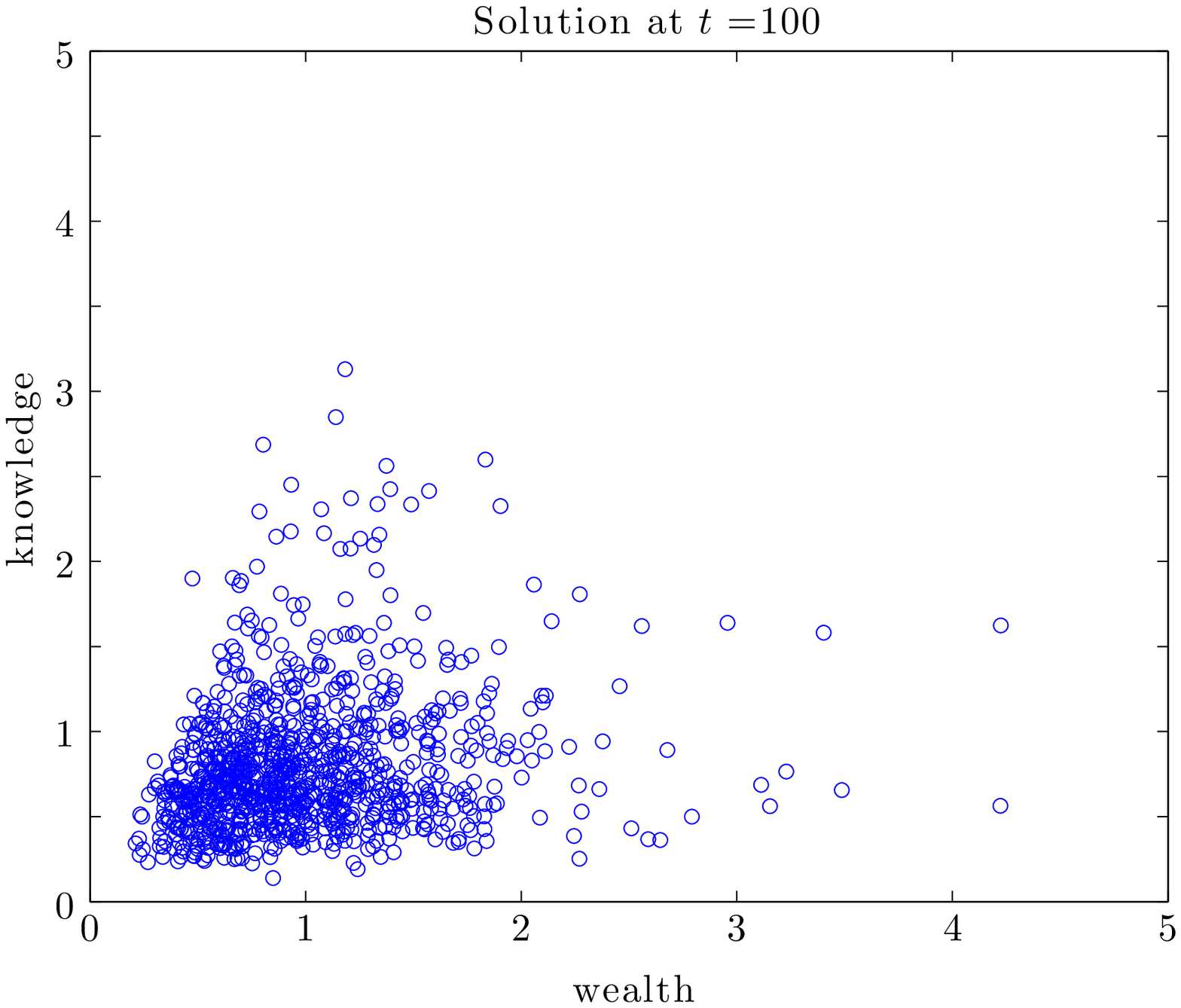}
\includegraphics[scale=.38]{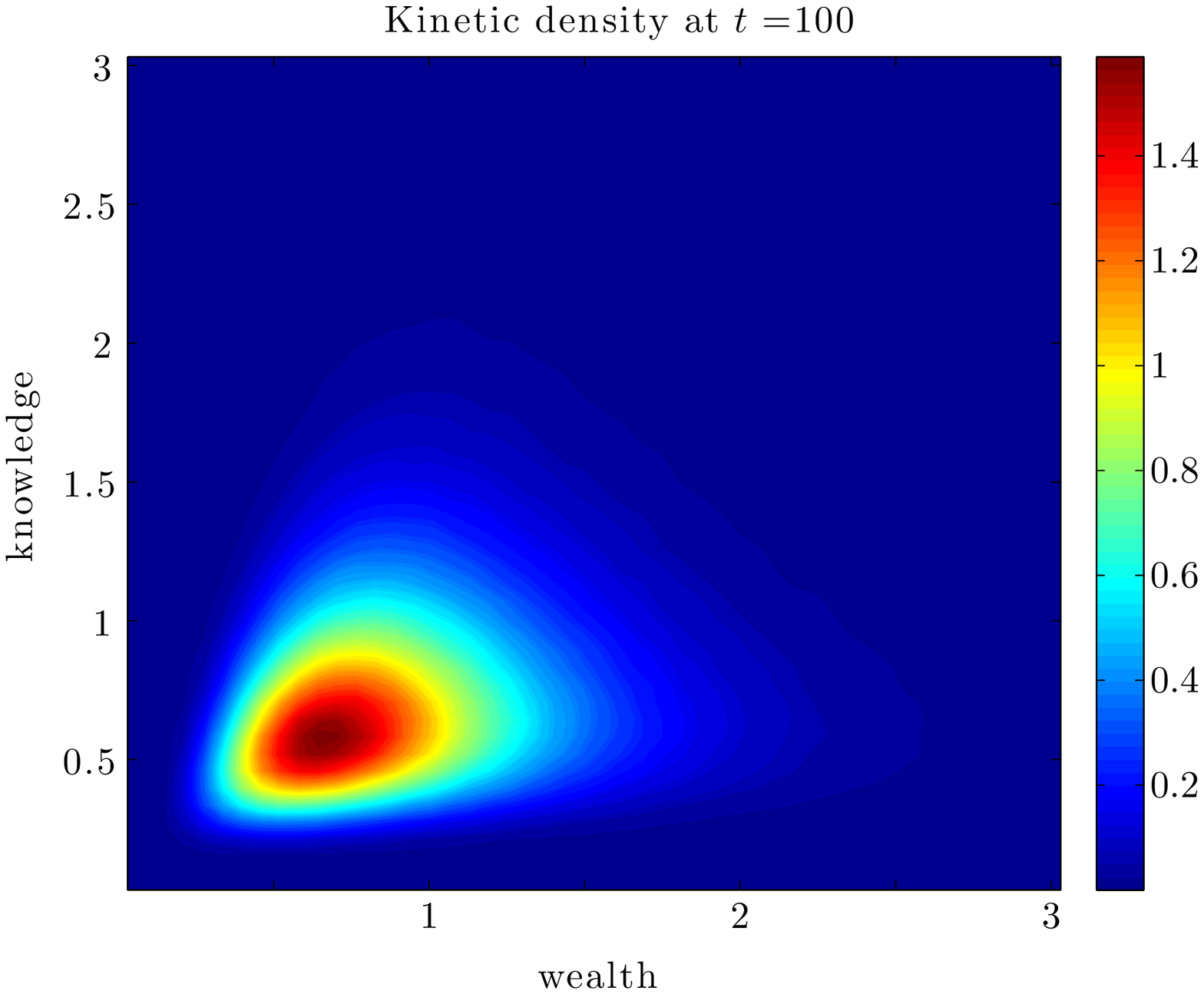}
\caption{Test 2: The particles solution with $N=1000$ agents (left) and the kinetic density (right).}
\label{fg:fig2}
\end{figure}

\subsubsection*{Test 1}
\begin{figure}[t]
\includegraphics[scale=0.38]{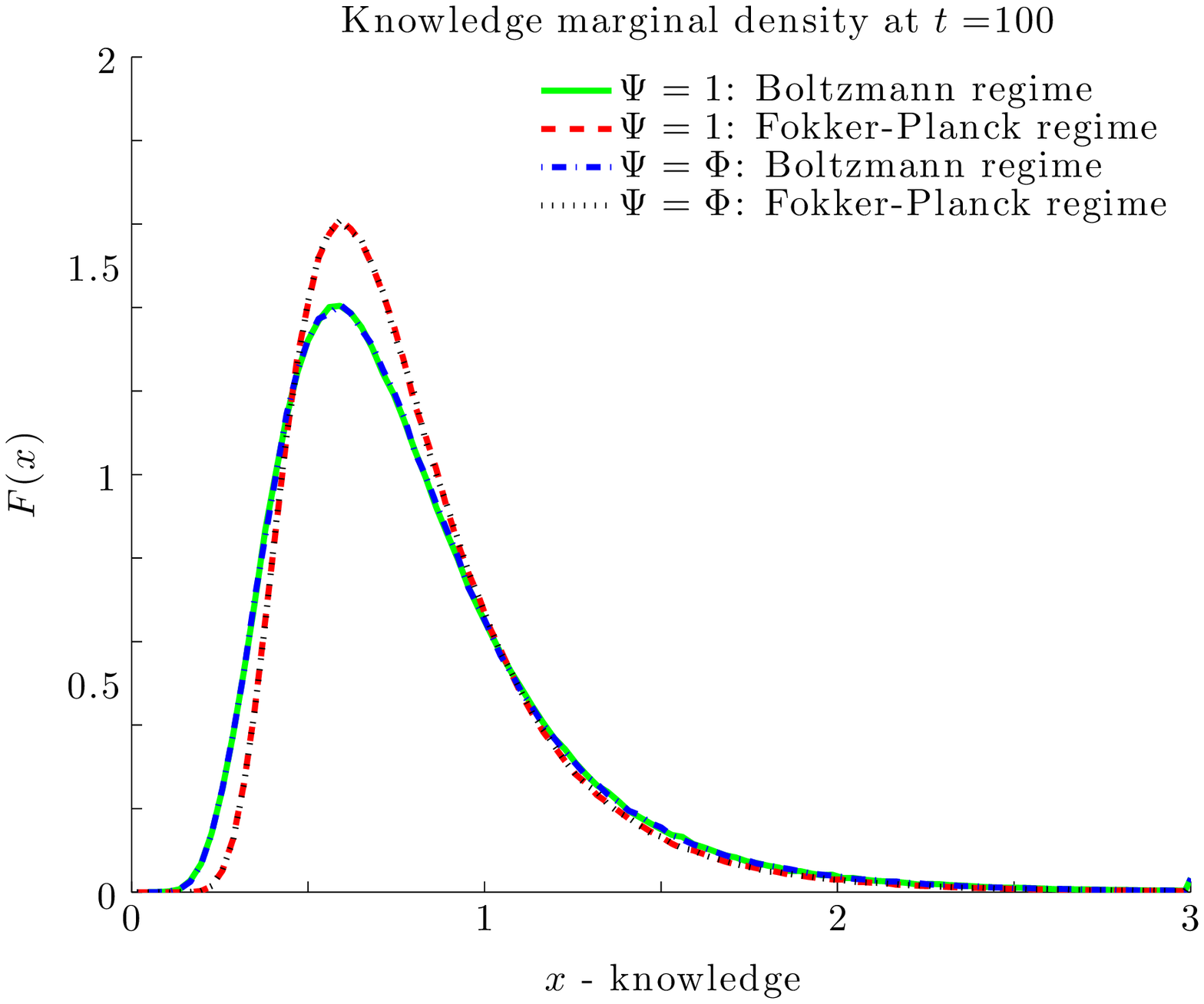}
\includegraphics[scale=0.38]{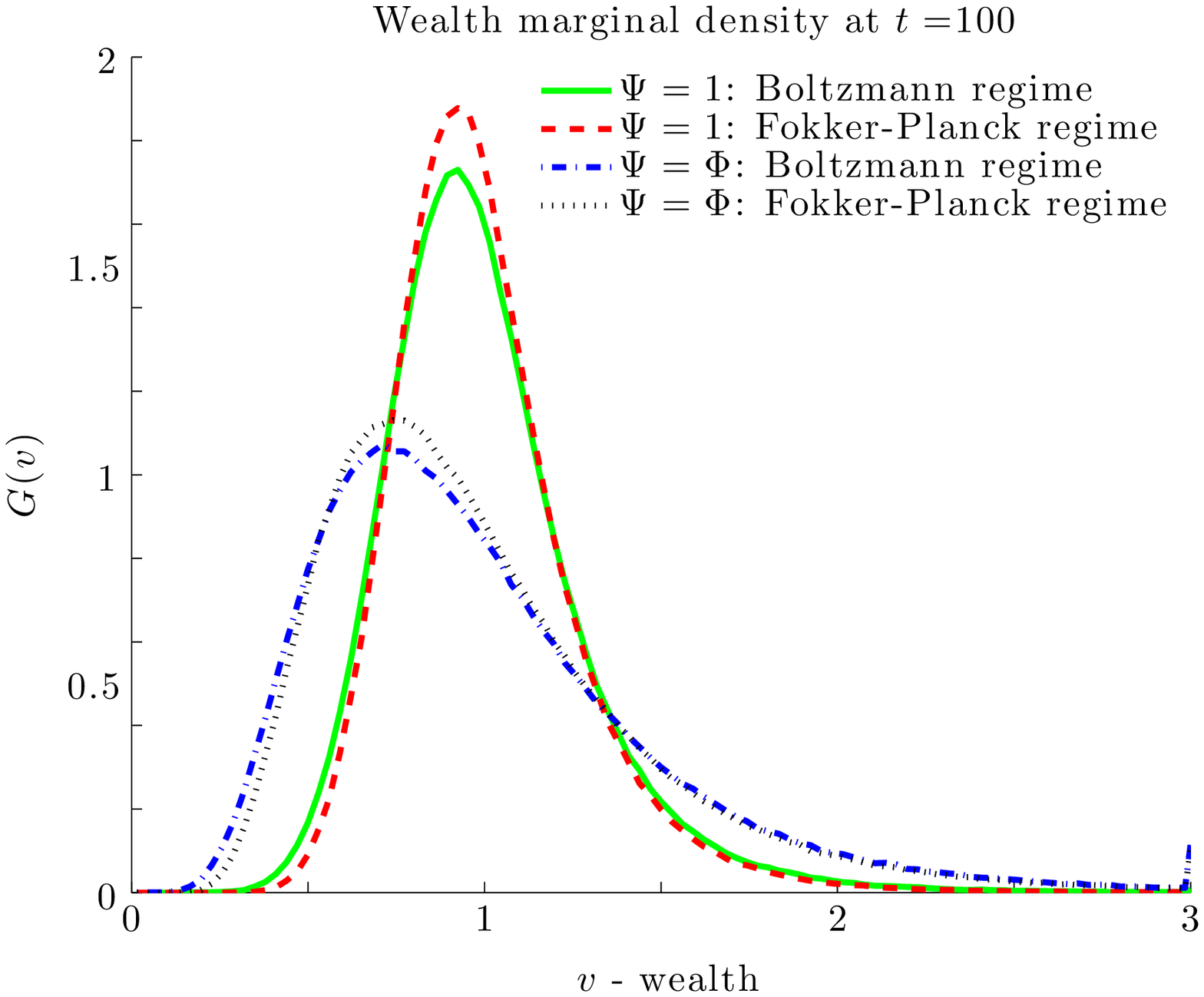}
\caption{The marginal densities in Test 1 and Test 2. Here $\lambda=\lambda_B=\gamma=\sigma=0.1$ in the Boltzmann regime and $\lambda=\lambda_B=\gamma=\sigma=0.01$ in the Fokker-Planck regime.}
\label{fg:fig3}
\end{figure}

\begin{figure}
\setlength{\unitlength}{1mm}
\begin{picture}(80,50)
\put(-2,0){
\hskip -.05cm
\includegraphics[scale=.38]{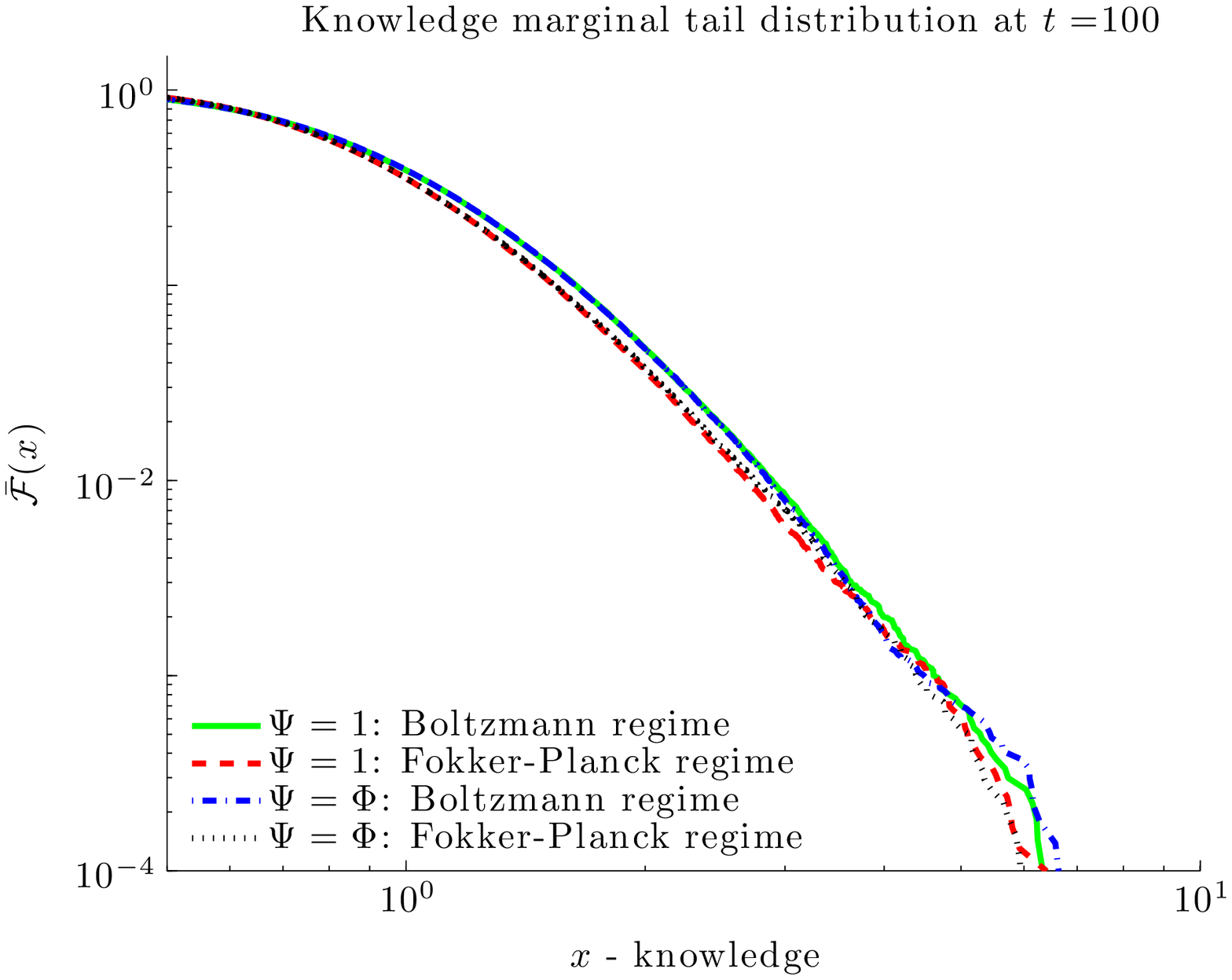}\hskip -.45cm
\includegraphics[scale=.38]{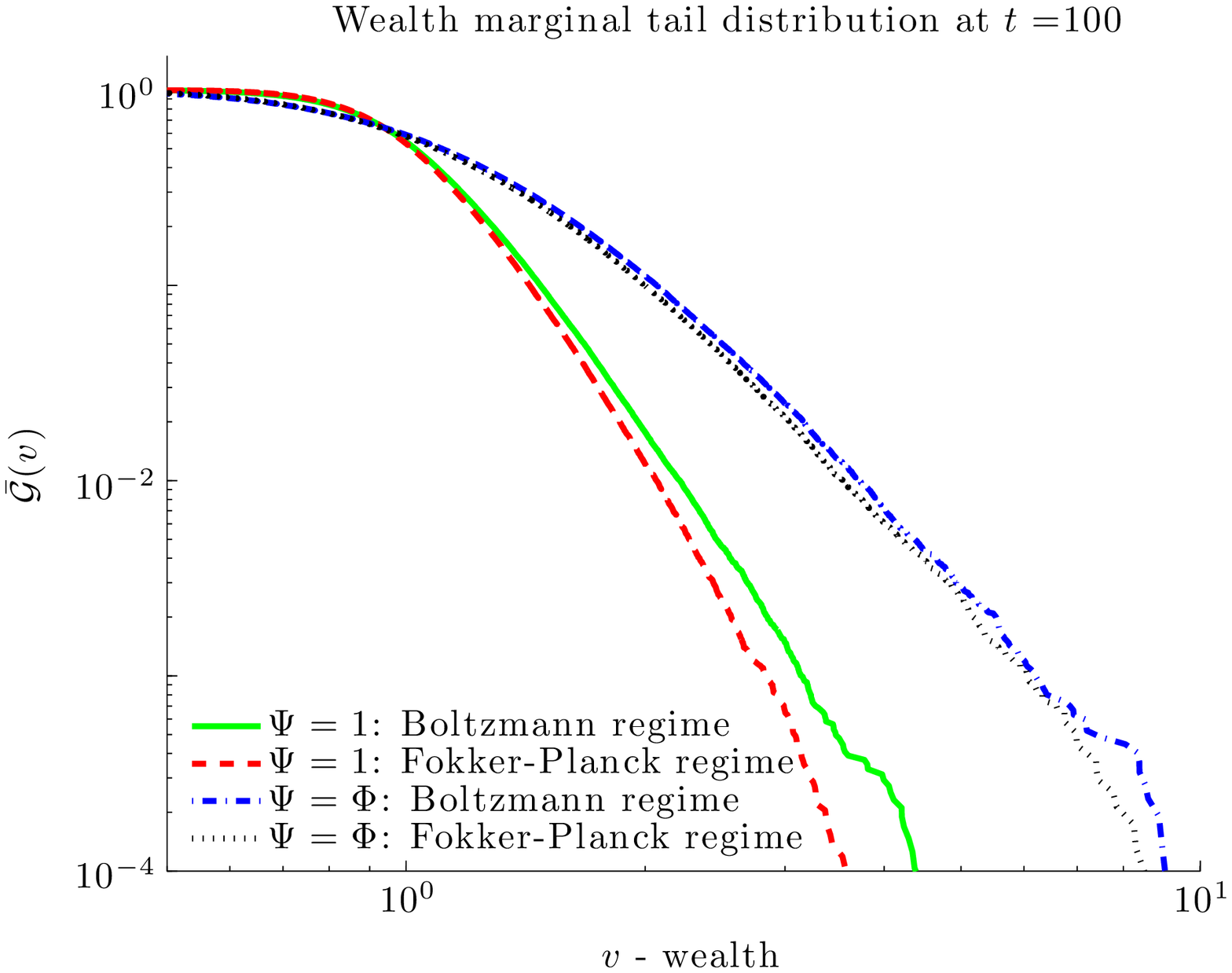}}
\put(51,20){\scriptsize $-4.7$}
\put(125,20){\scriptsize $-4.3$}
\put(111.5,20){\scriptsize $-6.2$}
\put(100,20){\scriptsize $-7.3$}
\put(51,23){\scriptsize slope}
\put(125,23){\scriptsize slope}
\put(111.5,23){\scriptsize slope}
\put(100,23){\scriptsize slope}
\end{picture}
\caption{The tail distribution of the marginal densities in Test 1 and Test 2 in log-log scale. Here $\lambda=\lambda_B=\gamma=\sigma=0.1$ in the Boltzmann regime and $\lambda=\lambda_B=\gamma=\sigma=0.01$ in the Fokker-Planck regime. The slope of the tail is estimated using least square fitting on the top $1\%$ of the agents population.}
\label{fg:fig4}
\end{figure}

In the first test we consider the case of a risk dependent knowledge with
\[
\Phi(x)=(1+x)^{-\beta},
\]
with $\beta=2$ and a constant $\Psi=1$. First we take $\lambda=\lambda_B=\gamma=\sigma=0.1$. These parameters correspond
to the Boltzmann regime. We consider a time step of $\Delta t=1$ and a final computation time of
$t=100$, where the steady state is practically reached. Since the
evolution of the knowledge in the model is independent from the
wealth, the latter is scaled in order to fix the mean wealth equal
to $1$. We report the results for the particle density and the
kinetic density in Figure \ref{fg:fig1}. In Figure \ref{fg:fig3} and \ref{fg:fig4} we
plot the marginal densities together with the tail distribution
\[
{\bar{\mathcal{F}}}(x)=1-\mathcal{F},\qquad {\bar{\mathcal{G}}}(x)=1-\mathcal{G},
\]
which are plotted in \emph{loglog} scale to visualize the tails
behavior. It is clear that the effect of the knowledge in minimizing
the risk produces a tendency for well educated people to occupy the
region around the mean wealth, the so-called \emph{middle class}. In
contrast the reach part of the people is composed by those who are
brave enough to take the risk in potentially remunerative
transactions.

The same test has been also performed in a Fokker-Planck
regime, scaling all interaction parameters by a factor $10$. The
final computation time is the same, but the time steps is now chosen
as $\Delta t=0.1$. The results in Fi\-gu\-res \ref{fg:fig3} and
\ref{fg:fig4} show a good agreement with the Boltzmann description.
As observed in \cite{CoPaTo05} the major differences are noticed in the peak of the distribution and in
the lower part of the population densities where the Fokker-Planck
solution is closer to zero. This latter effect is
due to the consistency of the boundary condition in $x=0$ and $v=0$ for the Fokker--Planck equation.


\subsubsection*{Test 2} In this new test, we maintain the same values for the parameters, and we modify the microscopic wealth interaction allowing for the knowledge to play a rule in the binary transaction. More precisely we assume that in \fer{eq.cpt2}
\[
\Psi(x)=\Phi(x).
\]
In this way the binary transaction is still conservative but a larger knowledge will correspond to a gain in the transaction. The results are reported in Figure \ref{fg:fig2} for the full density and in Figure \ref{fg:fig3} and \ref{fg:fig4} for the marginal densities and their tail distribution. It can be observed how this choice produces a shift of the whole distribution towards a more wealthy state for people with a higher knowledge. In particular the tails of the marginal density for the wealth are fatter if compared to the first test case. This shows that knowledge can play a relevant rule in the formation of wealth inequalities.

Lastly, for the above test cases in the Boltzmann regime, we report the profiles of the local mean wealth and local mean knowledge
\[
W(x,t)=\frac{1}{F(x,t)}\int f(x,v,t)v\,dv, \quad K(v,t)=\frac{1}{G(v,t)}\int f(x,v,t)x\,dx.
\]
The profiles clearly show that  the strategy of minimizing the risk is not enough to produce a class of well-educated and rich people. The main outcome is a concentration of agents with high values of knowledge around the mean wealth, but beside this the overall mean wealth for a given knowledge status is essentially independent from the knowledge.

\begin{figure}
\hskip -.1cm
\includegraphics[scale=.38]{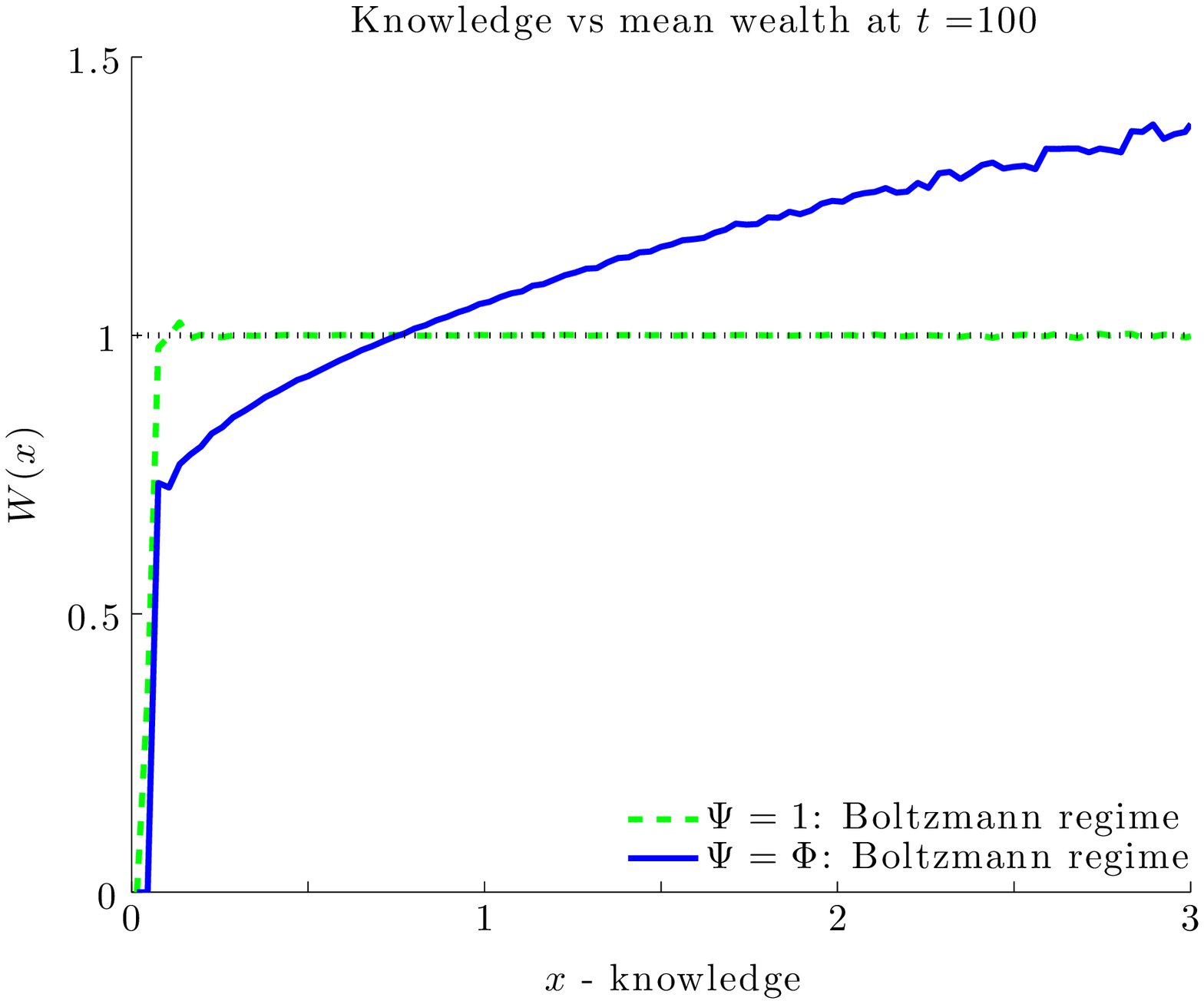}\hskip -.1cm
\includegraphics[scale=.38]{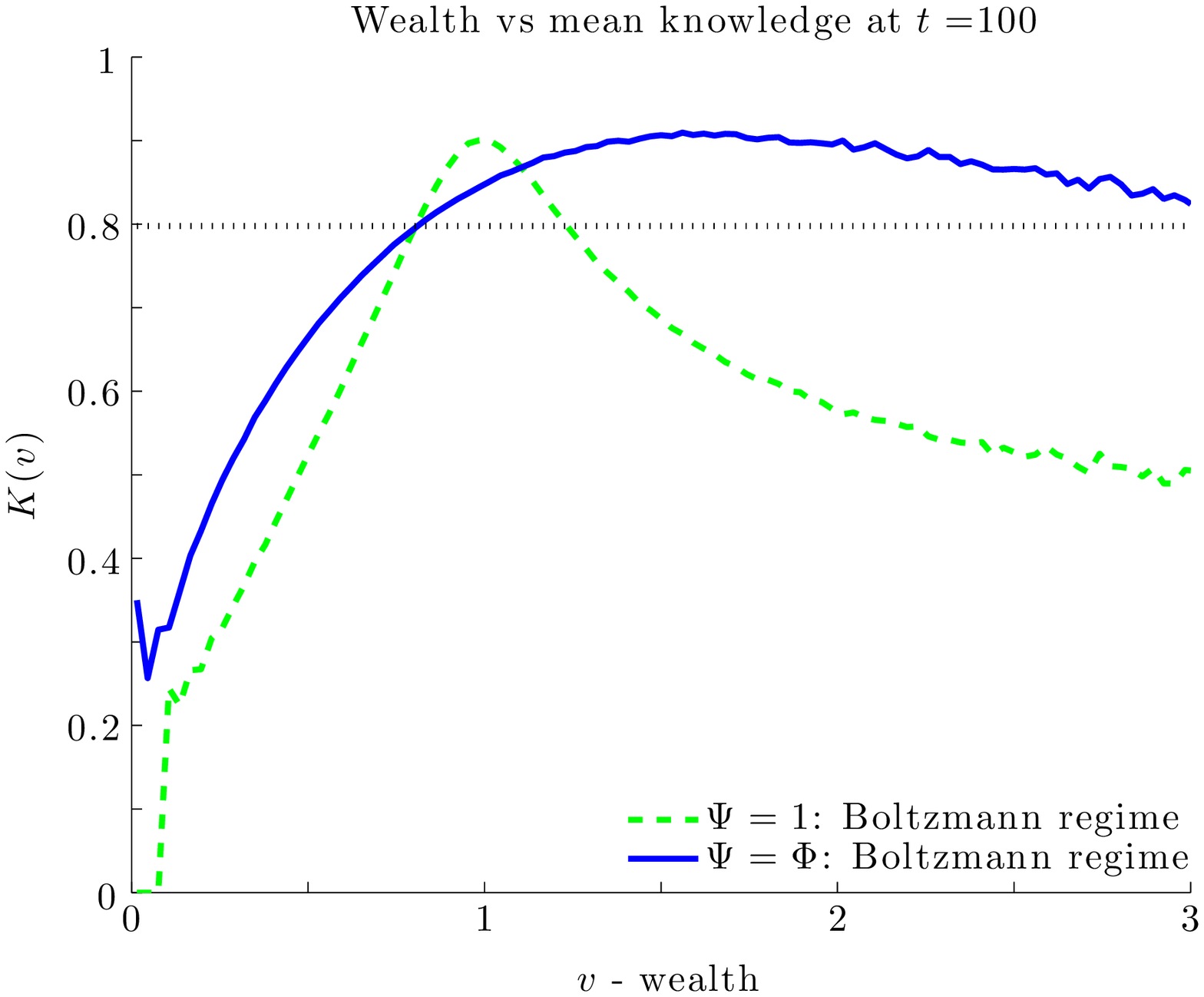}
\caption{The behavior of the local mean wealth (left) and local mean knowledge (right) in the case of Test 1 and Test 2. The dotted lines represent the mean wealth and mean knowledge.}
\label{fg:fig5}
\end{figure}

 On the contrary,  when the personal knowledge produces a certain advantage in the outcome of the binary trade, a shift is produced on the concentration of people with higher knowledge towards the richest classes and a strict correlation between wealth and knowledge is created (see Figure \ref{fg:fig5}).

\section{Conclusions}

Distribution of wealth in a society of agents depends on many aspects, even if it appears to have very stable features, like heavy tails, and  a relevant presence of the so-called middle class \cite{PT13}. In this note, we introduced and discussed a kinetic model for the joint evolution of wealth and knowledge, based on the assumption that knowledge can influence the distribution of wealth by acting on the risk perception, as well as in producing a better outcome from trades. Numerical experiments put in evidence that the role of knowledge in the distribution of wealth does not modify in a radical way the distribution of wealth, which remains heavy tailed in all the considered experiments. However, it is noticeable that the possibility to achieve a sensible richness strongly depends on the risky part of the trade. Hence, in all cases in which knowledge induces  to choose less-risky trades, the biggest part of the population which belongs to the middle class (with respect to knowledge) also end up with a location in the middle class (with respect to wealth). On the other hand when knowledge produces an advantage in the binary transaction it can account for strong wealth differences between the agents and a larger wealth inequality is originated.

\medskip
{\bf Acknowledgement.} This work has been done under the activities
of the National Group of Mathematical Physics (GNFM). The support of
the MIUR Research Projects of National Interest {\em ``Bayesian methods: theoretical developments and
novel applications''}, {\em ``Variational, functional-analytic, and
optimal transport methods for dissipative evolutions and stability
problems''} and {\em ``Advanced numerical methods for
kinetic equations and balance laws with source terms''} is kindly acknowledged.

\end{document}